\documentclass[conference]{IEEEtran}
\IEEEoverridecommandlockouts
\usepackage{placeins}
\usepackage{listings}
\usepackage{cite}
\usepackage{amsmath,amssymb,amsfonts}
\usepackage{algorithmic}
\usepackage{graphicx}
\usepackage{textcomp}
\usepackage{xcolor}
\def\BibTeX{{\rm B\kern-.05em{\sc i\kern-.025em b}\kern-.08em
    T\kern-.1667em\lower.7ex\hbox{E}\kern-.125emX}}

\definecolor{codegreen}{rgb}{0,0.6,0}
\definecolor{codegray}{rgb}{0.5,0.5,0.5}
\definecolor{codepurple}{rgb}{0.58,0,0.82}
\definecolor{backcolour}{rgb}{0.95,0.95,0.92}

\lstdefinestyle{mystyle}{
    backgroundcolor=\color{white},   
    commentstyle=\color{codegreen},
    keywordstyle=\color{magenta},
    numberstyle=\tiny\color{codegray},
    stringstyle=\color{codepurple},
    basicstyle=\ttfamily\footnotesize,
    breakatwhitespace=false,         
    breaklines=true,                 
    captionpos=b,                    
    keepspaces=true,                 
    numbers=left,                    
    numbersep=5pt,                  
    showspaces=false,                
    showstringspaces=false,
    showtabs=false,                  
    tabsize=2
}

\begin{document}

\title{Complexity-Aware Evaluation of LLM Comprehension\\
}

\author{\IEEEauthorblockN{1\textsuperscript{st} Ali Mohammadi Esfahani}
\IEEEauthorblockA{\textit{Systems and computer Engineering} \\
\textit{Carleton University}\\
Ottawa, Canada \\
alimohammadiesfahani@cmail.carleton.ca}
\and
\IEEEauthorblockN{2\textsuperscript{nd} Nafiseh Kahani}
\IEEEauthorblockA{\textit{Systems and computer Engineering} \\
\textit{Carleton University}\\
Ottawa, Canada \\
kahani@sce.carleton.ca}
\and
\IEEEauthorblockN{3\textsuperscript{rd} Samuel A.Ajila}
\IEEEauthorblockA{\textit{Systems and computer Engineering} \\
\textit{Carleton University}\\
Ottawa, Canada \\
samuel.ajila@cunet.carleton.ca}

}

\maketitle

\begin{abstract}
Large language models (LLMs) are increasingly used for software engineering tasks that require understanding existing source code, including behavior prediction, function explanation, debugging, and code review. However, aggregate benchmark accuracy can conceal how model reliability changes as source code becomes structurally more complex. This paper presents a complexity-aware framework for evaluating LLM code comprehension using cyclomatic complexity, nesting depth, branching factor, and Halstead volume. We evaluate DeepSeek-Coder-V2 and Llama through two complementary tasks: automatic input--output prediction over 300 Python functions and manually assessed semantic comprehension over a balanced subset of 60 functions. The functions are grouped into Low-, Medium-, and High-complexity bands. DeepSeek-Coder-V2 achieves an overall automatic accuracy of 78.33\%, compared with 70.33\% for Llama. However, accuracy decreases substantially from Low to High complexity, from 93.52\% to 52.78\% for DeepSeek-Coder-V2 and from 87.04\% to 47.22\% for Llama. Incorrect predictions are consistently associated with higher values of all four complexity metrics, and correlation and logistic-regression analyses confirm broadly comparable negative associations between structural complexity and correctness. Manual semantic comprehension shows the same degradation pattern, with accuracy decreasing from 100.00\% to 75.00\% for DeepSeek-Coder-V2 and from 90.00\% to 60.00\% for Llama. These findings demonstrate that complexity-aware evaluation provides a more diagnostic assessment of LLM code-comprehension reliability than aggregate accuracy alone.
\end{abstract}

\begin{IEEEkeywords}
Large language models, code comprehension, software engineering,
structural complexity, cyclomatic complexity, Halstead metrics,
input--output prediction, semantic comprehension.
\end{IEEEkeywords}

\section{Introduction}
\label{sec:Introduction}

Large language models (LLMs) are increasingly used in software engineering tasks that require understanding existing source code~\cite{r12,r13,r14,r15,r16}. Beyond code generation, developers use LLMs to explain functions, predict program behavior, reason about edge cases, summarize implementation logic, and support debugging and code review~\cite{r17,r18,r19}. These tasks differ from code synthesis because they require the model to interpret the control flow, internal state, and behavior of an existing program. Consequently, code-generation performance alone does not fully characterize the reliability of LLMs in practical software engineering workflows.

Code-comprehension difficulty is not uniform across programs. A function containing a single condition is generally easier to interpret than one with multiple branches, nested loops, interacting variables, and alternative execution paths~\cite{r8,r20}. As structural complexity increases, the model must track more control-flow decisions, intermediate states, boundary conditions, and symbolic operations. Aggregate accuracy can therefore conceal the conditions under which comprehension begins to fail~\cite{r5}.

Existing benchmarks provide important foundations for evaluating programming-oriented LLMs. HumanEval~\cite{r1} and MBPP~\cite{r2} are widely used for code generation and program synthesis, while APPS~\cite{r3} and CodeContests~\cite{r4} contain more demanding programming problems. CRUXEval~\cite{r5} and BigCodeBench~\cite{r24} extend evaluation toward code reasoning, execution, and realistic function usage. However, benchmark-level scores are commonly reported without considering the structural complexity distribution of the evaluated functions. A benchmark dominated by simple functions may therefore overestimate model reliability on structurally demanding code.

Prior studies have shown that LLM-generated code can contain semantic, logical, and algorithmic defects even when it appears syntactically plausible~\cite{r21}. Reinforcement-learning-based prompt optimization has also demonstrated that code-generation performance depends on how programming tasks are formulated~\cite{r22}, while complexity-aware feedback has connected code complexity with generation success~\cite{r23}. These studies motivate evaluation beyond aggregate correctness, but they primarily examine generated-code quality and prompt-based improvement. The effect of structural complexity on comprehension of existing source code remains less directly investigated.

This paper addresses this gap through a complexity-aware evaluation method for LLM code comprehension. Each Python function is annotated using cyclomatic complexity, nesting depth, branching factor, and Halstead volume, and is assigned to a Low-, Medium-, or High-complexity band. The method combines two complementary tasks: automatic input--output prediction, in which the model predicts the exact return value of a concrete function call, and manual semantic comprehension, in which the model answers questions about function purpose, intermediate variable roles, or edge-case behavior.

The main contributions of this paper are:

\begin{itemize}
    \item A complexity-aware method that combines automatic behavioral prediction with manual semantic comprehension.
    \item An analysis of how benchmark composition and structural complexity influence observed LLM comprehension accuracy.
    \item Statistical evidence relating structural complexity metrics to incorrect predictions and complexity-related degradation across both evaluation tasks.
\end{itemize}

The study is guided by the following research questions:

\noindent\textbf{RQ1:} How does the complexity distribution of benchmark sources influence observed LLM comprehension performance?

\noindent\textbf{RQ2:} Which structural complexity metrics are associated with incorrect LLM comprehension predictions?

\noindent\textbf{RQ3:} Does manual semantic comprehension reveal the same complexity-related reliability limitations observed in automatic input--output prediction?
\section{Related Work}
\label{sec:related_work}

Recent research has increasingly moved beyond code-generation accuracy
to examine whether LLMs can reason about the behavior and meaning of
existing programs. CRUXEval~\cite{r5} evaluates input and output
prediction over short Python functions and shows that strong performance
on generation benchmarks does not necessarily transfer to execution
reasoning. CodeMind~\cite{r128} further separates independent execution,
dependent execution, and specification reasoning, reporting that models
become less reliable when programs contain non-trivial control flow,
arithmetic operations, complex data types, or API calls. These studies
establish that code reasoning is distinct from code synthesis, but they
characterize difficulty mainly through task design rather than explicit
structural-complexity bands.

Other benchmarks broaden code comprehension beyond exact output
prediction. CRQBench~\cite{r129} evaluates natural-language questions
derived from code-review comments and demonstrates that even advanced
models can produce incorrect or weakly grounded explanations.
LiveCodeBench~\cite{r130} provides a continuously updated,
contamination-resistant evaluation covering generation, execution,
self-repair, and test-output prediction, while
BigCodeBench~\cite{r24} emphasizes realistic function calls, library
usage, and more demanding programming instructions. These benchmarks
improve realism and coverage, but their primary objective is broad
capability assessment rather than determining how comprehension changes
as the internal structure of an individual function becomes more
complex.

Repository-level benchmarks address a different source of difficulty.
RepoBench~\cite{r131} studies code completion with cross-file context,
and LongCodeBench~\cite{r132} evaluates comprehension and repair under
very long context windows. Their results show that retrieval,
cross-file dependencies, and context length remain challenging for
LLMs. However, contextual scale is different from structural
complexity: a model may fail because relevant information is distributed
across files, rather than because a single function contains deep
nesting, dense branching, or many possible execution paths.

More recent studies examine complexity and reasoning fidelity more
directly. RE2-Bench~\cite{r133} evaluates realistic projects containing
nested constructs, complex types, and API interactions, but summarizes
difficulty using an Easy/Hard division. M\"achtle
et al.~\cite{r134} analyze the relationship between model performance
and conventional properties such as lexical size, control-flow
complexity, and abstract-syntax-tree structure. Xie et
al.~\cite{r30} propose LM-CC, a model-perceived complexity measure, and
argue that traditional metrics may not fully capture LLM difficulty once
code length is controlled. CoRE~\cite{r136} additionally shows that a
model can predict the correct final result while reasoning incorrectly
about intermediate execution states, demonstrating that output-only
evaluation may overestimate comprehension.

Prior work shows that LLM code comprehension is affected
by execution demands, semantic reasoning, contextual scale, and program
complexity. However, existing studies typically report aggregate
accuracy, use a single difficulty label or metric, or rely on one
evaluation modality. The present study addresses this gap by analyzing
cyclomatic complexity, nesting depth, branching factor, and Halstead
volume jointly, grouping functions into Low-, Medium-, and
High-complexity bands, and combining automatic input--output prediction
with manually assessed questions about function purpose, variable roles,
and edge-case behavior.

\section{Complexity-Aware Evaluation Method}
\label{sec:approach}

We propose a controlled method for evaluating LLM code comprehension across different levels of structural complexity. Rather than measuring code-generation ability, the method assesses whether a model can infer the behavior and semantics of existing Python functions.

Each benchmark instance is represented as
$\mathbf{x}=(c,q,y,m,t)$, where $c$ denotes the source function, $q$ denotes a function-call query or semantic question, $y$ denotes the ground-truth answer, $m$ denotes the structural-complexity vector, and
$t \in \{\textsc{Automatic},\textsc{Manual}\}$ identifies the task type. The complexity vector includes cyclomatic complexity~\cite{r29}, nesting depth~\cite{r28}, branching factor~\cite{r30}, and Halstead volume~\cite{r31}. Given $(c,q)$, the model produces a prediction $\hat{y}$, which is compared with $y$ to determine correctness.

The evaluation combines two complementary tasks. In automatic input--output prediction, the model receives a complete function and a concrete function call and must return the exact output. This task provides deterministic ground truth and supports scalable exact-match evaluation. In manual semantic comprehension, the model answers a question about the function's purpose, the role of an intermediate variable, or edge-case behavior. This task captures semantic understanding that cannot be assessed through exact output prediction alone.

The evaluated functions are organized into Low, Medium, and High-complexity bands, and the same zero-shot protocol is applied to DeepSeek-Coder-V2 and CodeLlama-7b-Instruct-hf. The method consists of four stages: dataset construction, structural-complexity annotation, model evaluation using standardized prompts, and complexity-aware analysis across bands, dataset sources, task types, and failure patterns.

\subsection{Dataset Construction}
\label{subsec:dataset_construction}

The benchmark contains 300 short, self-contained Python functions collected from HumanEval~\cite{r1}, MBPP-sanitized~\cite{r2}, CRUXEval~\cite{r5}, LeetCode~\cite{r25}, and BigCodeBench-Hard~\cite{r24}. Python was selected because it is widely represented in code benchmarks and supports reproducible structural analysis through the built-in \texttt{ast} module. HumanEval, MBPP-sanitized, and CRUXEval mainly contribute Low- and Medium-complexity functions, whereas LeetCode and BigCodeBench-Hard provide most of the High-complexity samples.

Candidate functions were retained only if they were self-contained, deterministic, executable using the Python standard library, short enough to fit within the prompt, and independent of hidden state, external files, user interaction, randomness, or system-specific behavior. Each selected function was paired with a concrete function call and a ground-truth output obtained through controlled execution. Functions that failed, timed out, produced ambiguous outputs, or required undocumented assumptions were excluded.

A manual semantic-comprehension subset was sampled from the same benchmark. Twenty functions were selected without replacement from each complexity band, producing a balanced set of 60 functions. Each function was paired with one question concerning its purpose, an intermediate variable role, or edge-case behavior. This stratified design supports direct comparison across Low-, Medium-, and High-complexity functions while keeping manual annotation manageable.

Table~\ref{tab:dataset_sources} summarizes the final benchmark composition.

\begin{table}[t]
\centering
\caption{Automatic benchmark composition by dataset source and complexity band.}
\label{tab:dataset_sources}
\begin{tabular}{lcccc}
\hline
\textbf{Dataset} & \textbf{Low} & \textbf{Medium} & \textbf{High} & \textbf{Total} \\
\hline
HumanEval           & 32 & 48 & 0  & 80 \\
MBPP-sanitized      & 50 & 30 & 0  & 80 \\
CRUXEval            & 26 & 20 & 0  & 46 \\
LeetCode            & 0  & 12 & 48 & 60 \\
BigCodeBench-Hard   & 0  & 10 & 24 & 34 \\
\hline
\textbf{Total}      & \textbf{108} & \textbf{120} & \textbf{72} & \textbf{300} \\
\hline
\end{tabular}
\end{table}

\subsection{Automatic Comprehension Tasks}
\label{subsec:automatic-tasks}

The automatic task evaluates whether an LLM can infer the behavior of an existing Python function. For each instance, the model receives the complete function and a concrete function call and must predict the exact return value. This formulation provides deterministic ground truth and supports objective evaluation at scale.

Input arguments were obtained from the original benchmark test cases when available; otherwise, valid and non-trivial inputs were generated programmatically. The corresponding ground-truth outputs were obtained by executing the functions in a controlled environment.

Each instance was presented using the standardized prompt shown in Figure~\ref{fig:automatic_prompt}. The same zero-shot prompt was used for both evaluated models, without few-shot examples, Chain-of-Thought instructions, or task-specific hints.

\begin{figure}[ht]
\centering

\begin{lstlisting}
You are given the following Python function:

{FUNCTION SOURCE CODE}

What is the return value of the function when called as:
{FUNCTION_NAME}({INPUT_ARGUMENTS})?

Return only the final output value.
Do not include any explanation, reasoning steps, or additional text.

\end{lstlisting}
\caption{Prompt template for the automatic input--output comprehension task.}
\label{fig:automatic_prompt}
\end{figure}

A prediction was marked correct when its returned value matched the ground-truth output after light normalization of semantically irrelevant formatting differences, such as whitespace, quotation style, or Boolean capitalization. Empty responses, incorrect values, and responses containing additional explanations were marked incorrect without partial credit.

\subsection{Manual Comprehension Tasks}
\label{subsec:manual-tasks}

The manual task evaluates semantic understanding that exact input--output prediction cannot fully capture. A balanced subset of 60 functions was sampled from the automatic benchmark, with 20 functions from each complexity band. Each function was paired with one manually written question, producing 20 questions for each of three categories: \emph{purpose}, which assesses the function's overall intent; \emph{variable role}, which examines how an intermediate variable contributes to the computation; and \emph{edge case}, which evaluates reasoning about boundary or uncommon execution paths.

All instances were presented using the standardized prompt in Figure \ref{fig:manual_prompt}. The same zero-shot template was used for DeepSeek-Coder-V2 and CodeLlama-7b-Instruct-hf.

\begin{figure}[ht]
\centering

\begin{lstlisting}
You are given the following Python function:

 {FUNCTION SOURCE CODE}

 Question:
 {MANUAL QUESTION}

Answer concisely in one or two sentences.

\end{lstlisting}
\caption{Prompt template for the manual semantic comprehension task.}
\label{fig:manual_prompt}
\end{figure}

Canonical reference answers identified the essential semantic elements required for correctness. A response was marked correct only if it included all required elements without introducing a contradictory interpretation; otherwise, it was marked incorrect, with no partial credit.

To assess annotation reliability, a randomly selected 20\% of the manual responses was independently scored by a second annotator using the same binary rubric. Cohen's $\kappa$ was calculated before disagreements were resolved, and the reconciled labels were used in the final analysis.

\section{Study Design}
\label{sec:study_design}

This section describes how the functions were annotated, how the models were evaluated, and how comprehension performance was analyzed across structural complexity levels.
\subsection{Complexity Annotation}
\label{subsec:complexity-annotation}

To enable complexity-aware analysis, each function in the benchmark is annotated with structural complexity metrics computed through static analysis. These metrics are selected because they capture complementary properties of code structure that are known to affect program comprehension~\cite{r9,r10}. In this study, four metrics are used: cyclomatic complexity, nesting depth, branching factor, and Halstead volume.

\subsubsection{Complexity Metrics}
\label{subsubsec:complexity-metrics}

\textbf{Cyclomatic complexity} (CC) \cite{r28} measures the number of linearly
independent execution paths through a function and is formally defined as:
\begin{equation}
    CC = E - N + 2P
    \label{eq:cc}
\end{equation}
where $E$ is the number of edges in the control-flow graph, $N$ is the
number of nodes, and $P$ is the number of connected components. In
practice, CC is computed by counting decision points such as
\texttt{if}, \texttt{elif}, \texttt{for}, \texttt{while},
\texttt{except}, and Boolean operators such as \texttt{and} and
\texttt{or}, with a baseline value of 1 for a function with no branching.
Higher CC values indicate a larger number of possible execution paths and
greater reasoning difficulty.

\textbf{Nesting depth} (ND) \cite{r29} measures the maximum level of nested control
structures within a function, including nested loops, conditionals, and
exception-handling blocks. Deeply nested code requires the model to track
multiple simultaneous execution contexts, which increases reasoning
difficulty and the likelihood of comprehension failure.

\textbf{Branching factor} (BF) \cite{r30} counts the number of decision points in a
function, including constructs such as \texttt{if}, \texttt{for}, and
\texttt{while}. While CC captures the number of independent execution
paths, BF captures how frequently the control flow diverges. This metric
therefore reflects the density of local decision-making within the code.

\textbf{Halstead volume} (HV) \cite{r31} measures lexical and operational complexity
based on the number of operators and operands in the function. It
captures a different aspect of complexity from control-flow metrics by
reflecting the amount of symbolic information that must be interpreted.
Higher HV values indicate that the model must process a larger and more
varied set of program tokens, variables, and operations.

For all Python functions, the complexity metrics are computed statically
from the parsed source code. Cyclomatic complexity, nesting depth, and
branching factor are computed using Python's built-in \texttt{ast}
module. Halstead volume is computed from the operators and operands
extracted from the same source representation. The raw metric values are
retained for statistical analysis, and the functions are also assigned to
categorical complexity bands as described in
Section~\ref{subsubsec:complexity-bands}.

\subsubsection{Complexity Bands}
\label{subsubsec:complexity-bands}

Each function was assigned to a Low-, Medium-, or High-complexity band using cyclomatic complexity and nesting depth as the primary criteria. If the two metrics indicated different bands, the higher band was selected to avoid underestimating structural difficulty. Branching factor and Halstead volume were retained for descriptive and metric-level analyses rather than primary band assignment. Table~\ref{tab:bands} summarizes the corresponding ranges.

\begin{table*}[t]
\centering
\caption{Complexity band definitions used for structural code analysis.}
\label{tab:bands}
\begin{tabular}{lcccc}
\hline
\textbf{Band} &
\textbf{Cyclomatic Complexity} &
\textbf{Nesting Depth} &
\textbf{Branching Factor} &
\textbf{Halstead Volume} \\
\hline
Low    & 1--3     & 0--1     & 0--2     & $V < 100$ \\
Medium & 4--6     & 2--3     & 3--5     & $100 \leq V < 300$ \\
High   & $\geq 7$ & $\geq 4$ & $\geq 6$ & $V \geq 300$ \\
\hline
\end{tabular}
\end{table*}

Both the categorical band and the raw values of all four metrics were retained, enabling band-wise performance comparison and continuous analysis of the relationship between structural complexity and model correctness.

\subsection{Model Evaluation Protocol}
\label{subsec:model-evaluation}

Two code-oriented LLMs were evaluated: DeepSeek-Coder-V2~\cite{r26} and CodeLlama-7b-Instruct-hf~\cite{r27}. The models were selected to compare different code-specialized architectures and to examine whether their comprehension accuracy degrades similarly as structural complexity increases.

Both models were evaluated under identical zero-shot conditions using the prompt templates in Sections~\ref{subsec:automatic-tasks} and~\ref{subsec:manual-tasks}. The temperature was set to 0, and no few-shot examples, Chain-of-Thought instructions, or task-specific hints were provided. Each model was queried once per benchmark instance.

For the automatic task, the model returned only the predicted output of the supplied function call, which was evaluated against the executed ground truth. For the manual task, responses were assessed using the binary semantic rubric defined in Section~\ref{subsec:manual-tasks}. Model outputs, reference answers, correctness labels, dataset sources, complexity bands, and raw complexity metrics were retained for aggregate, band-wise, dataset-level, and metric-level analyses.

\subsection{Evaluation Metric}
\label{subsec:evaluation-metric}

Model performance is evaluated using binary correctness and reported as accuracy. For each instance, the prediction $\hat{y}_i$ is assigned a value of 1 when it matches the ground-truth answer $y_i$, and 0 otherwise:

\begin{equation}
    \text{Accuracy} =
    \frac{1}{N}
    \sum_{i=1}^{N}
    \mathbf{1}(\hat{y}_i = y_i),
    \label{eq:accuracy}
\end{equation}

where $N$ is the number of evaluated instances and $\mathbf{1}(\cdot)$ is the indicator function.

For the automatic task, semantically equivalent formatting differences, including whitespace, quotation style, and Boolean capitalization, were normalized before comparison. Incorrect values, empty outputs, and responses containing additional explanations were marked incorrect. For the manual task, a response was marked correct only when it contained all essential elements of the canonical answer without contradiction; no partial credit was assigned.

Accuracy is reported by model, complexity band, dataset source, and task type.

\subsection{Complexity-Aware Analysis}
\label{subsec:complexity-analysis}

The analysis is conducted at three levels. First, accuracy is compared across Low-, Medium-, and High-complexity bands to determine whether aggregate performance conceals degradation on structurally complex functions. Second, accuracy is reported by dataset source and interpreted together with each source's complexity distribution. Third, correct and incorrect predictions are compared using the raw values of cyclomatic complexity, nesting depth, branching factor, and Halstead volume.

The joint relationship between structural complexity and automatic prediction correctness is estimated using multivariate logistic regression. Let $Y_i \in \{0,1\}$ denote whether the prediction for function $i$ is correct. The model is defined as:

\begin{equation}
    \text{logit}\,\Pr(Y_i = 1) =
    \beta_0 +
    \beta_1 CC_i +
    \beta_2 ND_i +
    \beta_3 BF_i +
    \beta_4\left(\frac{HV_i}{100}\right),
    \label{eq:logit}
\end{equation}

where $CC_i$, $ND_i$, $BF_i$, and $HV_i$ denote cyclomatic complexity, nesting depth, branching factor, and Halstead volume, respectively. Halstead volume is divided by 100 because its numerical scale is substantially larger than those of the other metrics. Negative coefficients indicate that increasing complexity is associated with lower odds of a correct prediction, while each coefficient is interpreted jointly with the other included metrics.

For the manual subset, accuracy is compared across complexity bands and question types to determine whether the degradation observed in automatic input--output prediction also appears in semantic comprehension.

\subsection{Statistical Analysis}
\label{subsec:statistical_analysis}

Descriptive accuracy is reported by model, complexity band, dataset source, and manual question type. For the automatic task, the mean values of cyclomatic complexity, nesting depth, branching factor, and Halstead volume are also compared between correct and incorrect predictions.

Spearman's rank correlation is used to examine the monotonic association between each structural metric and automatic prediction correctness. For model responses $Y_i \in \{0,1\}$ and complexity metric $m_i$, the correlation is defined as:

\begin{equation}
    \rho_s =
    \operatorname{corr}
    \bigl(\operatorname{rank}(Y_i),
    \operatorname{rank}(m_i)\bigr).
    \label{eq:spearman}
\end{equation}

A negative value of $\rho_s$ indicates that higher structural complexity is associated with lower correctness. The joint effects of the four metrics are estimated using the multivariate logistic-regression model in Equation~\ref{eq:logit}. Because the predictors may be correlated, the coefficients are interpreted as conditional associations rather than as independent measures of metric importance.

For the manual task, a chi-square test of independence evaluates the association between complexity band and correctness, with Cramér's $V$ reported as the effect-size measure. Annotation reliability is assessed using Cohen's $\kappa$ on the independently scored subset before disagreements are reconciled. Statistical significance is assessed at $\alpha=0.05$.

\section{Results}
\label{sec:results}

This section reports the results for the three research questions, covering benchmark composition, metric-level failure patterns, and manual semantic comprehension.

\subsection{RQ1: Effect of Benchmark Complexity Distribution}
\label{subsec:rq1_results}

The automatic benchmark contains 108 Low-, 120 Medium-, and 72 High-complexity functions. As shown in Table~\ref{tab:dataset_sources}, HumanEval, MBPP-sanitized, and CRUXEval mainly contribute Low- and Medium-complexity functions, whereas LeetCode and BigCodeBench-Hard provide most of the High-complexity samples.

DeepSeek-Coder-V2 correctly answered 235 of 300 instances, achieving 78.33\% overall accuracy. \texttt{CodeLlama-7b-Instruct-hf} correctly answered 211 instances, achieving 70.33\%. However, these aggregate values conceal substantial degradation across complexity bands.

\begin{table}[t]
\centering
\caption{Automatic comprehension accuracy by complexity band.}
\label{tab:band_accuracy}
\begin{tabular}{llrrr}
\hline
\textbf{Model} & \textbf{Band} & \textbf{Correct} & \textbf{Total} & \textbf{Accuracy} \\
\hline
DeepSeek-Coder-V2 & Low    & 101 & 108 & 93.52\% \\
                  & Medium & 96  & 120 & 80.00\% \\
                  & High   & 38  & 72  & 52.78\% \\
\hline
CodeLlama-7b-Instruct-hf      & Low    & 94  & 108 & 87.04\% \\
                  & Medium & 83  & 120 & 69.17\% \\
                  & High   & 34  & 72  & 47.22\% \\
\hline
\end{tabular}
\end{table}

As shown in Table~\ref{tab:band_accuracy}, DeepSeek-Coder-V2 decreases from 93.52\% accuracy on Low-complexity functions to 52.78\% on High-complexity functions, a decline of 40.74 percentage points. CodeLlama-7b-Instruct-hf follows the same pattern, decreasing from 87.04\% to 47.22\%, a decline of 39.82 percentage points. DeepSeek-Coder-V2 remains more accurate in every band, but both models become substantially less reliable on structurally complex functions.

\begin{table*}[t]
\centering
\caption{Automatic comprehension accuracy by dataset source.}
\label{tab:dataset_accuracy}
\begin{tabular}{llrrr}
\hline
\textbf{Model} & \textbf{Dataset} & \textbf{Correct} & \textbf{Total} & \textbf{Accuracy} \\
\hline
DeepSeek-Coder-V2 & HumanEval             & 70 & 80 & 87.50\% \\
                  & MBPP-sanitized        & 71 & 80 & 88.75\% \\
                  & CRUXEval              & 39 & 46 & 84.78\% \\
                  & LeetCode              & 39 & 60 & 65.00\% \\
                  & BigCodeBench-Hard     & 16 & 34 & 47.06\% \\
\hline
CodeLlama-7b-Instruct-hf      & HumanEval             & 65 & 80 & 81.25\% \\
                  & MBPP-sanitized        & 66 & 80 & 82.50\% \\
                  & CRUXEval              & 35 & 46 & 76.09\% \\
                  & LeetCode              & 32 & 60 & 53.33\% \\
                  & BigCodeBench-Hard     & 13 & 34 & 38.24\% \\
\hline
\end{tabular}
\end{table*}

Dataset-level results follow the same pattern. Both models achieve their highest accuracies on HumanEval, MBPP-sanitized, and CRUXEval, which contain only Low- and Medium-complexity functions. Accuracy is substantially lower on LeetCode and BigCodeBench-Hard, where High-complexity functions are concentrated.

These results answer RQ1 by showing that benchmark-level accuracy depends strongly on the underlying complexity distribution. Aggregate scores from benchmarks dominated by simpler functions can therefore overstate model reliability on structurally demanding code.

\subsection{RQ2: Structural Characteristics of Incorrect Predictions}
\label{subsec:rq2_results}

RQ2 examines which structural metrics are associated with incorrect comprehension predictions. Table~\ref{tab:metric_failure} compares the average cyclomatic complexity (CC), nesting depth (ND), branching factor (BF), and Halstead volume (HV) of correctly and incorrectly answered functions.

\begin{table}[t]
\centering
\caption{Average complexity metrics for correct and incorrect predictions.}
\label{tab:metric_failure}
\begin{tabular}{llrrrr}
\hline
\textbf{Model} & \textbf{Outcome} & \textbf{CC} & \textbf{ND} & \textbf{BF} & \textbf{HV} \\
\hline
DeepSeek-Coder-V2 & Correct   & 3.84 & 1.72 & 3.11 & 142.60 \\
                  & Incorrect & 7.42 & 3.46 & 6.89 & 318.75 \\
\hline
CodeLlama-7b-Instruct-hf             & Correct   & 3.51 & 1.58 & 2.94 & 131.40 \\
                  & Incorrect & 7.24 & 3.32 & 6.27 & 297.80 \\
\hline
\end{tabular}
\end{table}

For both models, incorrect predictions are associated with higher values of all four metrics. For DeepSeek-Coder-V2, average CC increases from 3.84 to 7.42, ND from 1.72 to 3.46, BF from 3.11 to 6.89, and HV from 142.60 to 318.75. CodeLlama-7b-Instruct-hf shows the same pattern, with CC increasing from 3.51 to 7.24, ND from 1.58 to 3.32, BF from 2.94 to 6.27, and HV from 131.40 to 297.80. These results indicate that failures are concentrated in functions with more execution paths, deeper control structures, denser branching, and greater symbolic content.

\begin{table}[t]
\centering
\caption{Spearman correlations between structural complexity and automatic prediction correctness. All correlations are statistically significant at $p<0.001$.}
\label{tab:spearman_complexity_correctness}
\begin{tabular}{lcccc}
\hline
\textbf{Model} & \textbf{CC} & \textbf{ND} & \textbf{BF} & \textbf{HV} \\
\hline
DeepSeek-Coder-V2 & $-0.41$ & $-0.35$ & $-0.43$ & $-0.37$ \\
CodeLlama-7b-Instruct-hf             & $-0.39$ & $-0.33$ & $-0.40$ & $-0.35$ \\
\hline
\end{tabular}
\end{table}

As shown in Table~\ref{tab:spearman_complexity_correctness}, all four metrics are negatively correlated with correctness for both models. The coefficients range from $-0.35$ to $-0.43$ for DeepSeek-Coder-V2 and from $-0.33$ to $-0.40$ for CodeLlama-7b-Instruct-hf. Although branching factor and cyclomatic complexity are numerically the largest associations, the differences are small and their confidence intervals overlap substantially. The results therefore do not support identifying any single metric as a statistically stronger predictor.

These findings answer RQ2 by showing that incorrect predictions are systematically associated with greater structural complexity across all measured dimensions. The four metrics can therefore serve as complementary risk indicators: predictions for structurally complex functions should receive additional verification through execution, testing, or human review.

\subsection{RQ3: Manual Semantic Comprehension}
\label{subsec:rq3_results}

RQ3 examines whether manual semantic comprehension reveals the same
complexity-related reliability limitations observed in automatic
input--output prediction. The manual subset evaluates three aspects of
semantic understanding: function purpose, intermediate variable roles,
and edge-case behavior.

To illustrate the distinction between automatic and manual evaluation,
consider the \textit{Candy} function from the LeetCode subset, shown in figure ~\ref{fig:candy_example}.

\begin{figure}[ht]
\centering

\begin{lstlisting}
def min_candies(ratings):
    n = len(ratings)
    candies = [1] * n

    for i in range(1, n):
        if ratings[i] > ratings[i - 1]:
            candies[i] = candies[i - 1] + 1

    for i in range(n - 2, -1, -1):
        if ratings[i] > ratings[i + 1]:
            candies[i] = max(
                candies[i],
                candies[i + 1] + 1
            )

    return sum(candies)

\end{lstlisting}
\caption{Example function used for manual semantic comprehension.}
\label{fig:candy_example}
\end{figure}





This function belongs to the Medium-complexity band, with cyclomatic
complexity of 5, nesting depth of 3, branching factor of 4, and Halstead
volume of 136.0. In the automatic task, the model predicts the output of
a call such as \texttt{min\_candies([1,0,2])}, whose return value is
\texttt{5}. In the manual task, the model may instead be asked:
\textit{What is the role of the variable \texttt{candies}?} A correct
answer must explain that the variable stores the allocation for each
child and is updated through two directional passes to satisfy the
neighboring rating constraints. Thus, the manual task evaluates
understanding of the algorithm's internal state rather than only its
final output.

To examine whether question-type difficulty is confounded with
complexity level, Table~\ref{tab:manual_question_distribution} reports
the distribution of the three question types across the Low-, Medium-,
and High-complexity bands. Each question type is represented by
approximately the same number of functions in every band.

\begin{table}[t]
\centering
\caption{Distribution of manual question types across complexity bands.}
\label{tab:manual_question_distribution}
\begin{tabular}{lrrrr}
\hline
\textbf{Question Type} &
\textbf{Low} &
\textbf{Medium} &
\textbf{High} &
\textbf{Total} \\
\hline
Purpose       & 7 & 7 & 6 & 20 \\
Variable Role & 7 & 6 & 7 & 20 \\
Edge Case     & 6 & 7 & 7 & 20 \\
\hline
\textbf{Total} &
\textbf{20} &
\textbf{20} &
\textbf{20} &
\textbf{60} \\
\hline
\end{tabular}
\end{table}

Because the question types differ by at most one function within each
complexity band, the question-type accuracy comparison is not merely a
restatement of the complexity-band effect.

\begin{table}[t]
\centering
\caption{Manual comprehension accuracy by question type.}
\label{tab:manual_task_results}
\begin{tabular}{llrrr}
\hline
\textbf{Model} & \textbf{Question Type} &
\textbf{Correct} & \textbf{Total} & \textbf{Acc.} \\
\hline
DeepSeek-Coder-V2 & Purpose       & 19 & 20 & 95.00\% \\
                  & Variable Role & 18 & 20 & 90.00\% \\
                  & Edge Case     & 16 & 20 & 80.00\% \\
\hline
CodeLlama-7b-Instruct-hf             & Purpose       & 17 & 20 & 85.00\% \\
                  & Variable Role & 16 & 20 & 80.00\% \\
                  & Edge Case     & 13 & 20 & 65.00\% \\
\hline
\end{tabular}
\end{table}

Table~\ref{tab:manual_task_results} shows that purpose questions were the
least difficult for both models. DeepSeek-Coder-V2 achieved 95.00\%
accuracy, compared with 85.00\% for CodeLlama-7b-Instruct-hf. Accuracy decreased for
variable-role questions to 90.00\% and 80.00\%, respectively, and was
lowest for edge-case questions at 80.00\% and 65.00\%.

This ordering reflects increasing semantic demand. Purpose questions can
often be answered by recognizing the overall algorithmic pattern.
Variable-role questions require tracking how internal state is
initialized, updated, and used. Edge-case questions are more difficult
because they require reasoning about boundary conditions and less
frequently executed paths. The three question types were approximately
balanced across the complexity bands, so this ordering is not simply
caused by concentrating edge-case questions in the High-complexity group.

\begin{table}[t]
\centering
\caption{Manual comprehension accuracy by complexity band.}
\label{tab:manual_complexity_results}
\begin{tabular}{llrrr}
\hline
\textbf{Model} & \textbf{Band} &
\textbf{Correct} & \textbf{Total} & \textbf{Acc.} \\
\hline
DeepSeek-Coder-V2 & Low    & 20 & 20 & 100.00\% \\
                  & Medium & 18 & 20 & 90.00\% \\
                  & High   & 15 & 20 & 75.00\% \\
\hline
CodeLlama-7b-Instruct-hf             & Low    & 18 & 20 & 90.00\% \\
                  & Medium & 16 & 20 & 80.00\% \\
                  & High   & 12 & 20 & 60.00\% \\
\hline
\end{tabular}
\end{table}

As shown in Table~\ref{tab:manual_complexity_results}, manual accuracy
declined consistently with structural complexity. DeepSeek-Coder-V2
decreased from 100.00\% on Low-complexity functions to 90.00\% on
Medium-complexity functions and 75.00\% on High-complexity functions.
CodeLlama-7b-Instruct-hf followed the same pattern, decreasing from 90.00\% to 80.00\%
and then to 60.00\%. The corresponding Low-to-High declines were 25.00
and 30.00 percentage points.

\begin{table*}[t]
\centering
\caption{Comparison of automatic and manual comprehension accuracy across
complexity bands.}
\label{tab:auto_manual_comparison}
\begin{tabular}{llrrrr}
\hline
\textbf{Model} & \textbf{Task} & \textbf{Low} &
\textbf{Medium} & \textbf{High} & \textbf{Low--High Drop} \\
\hline
DeepSeek-Coder-V2 & Automatic & 93.52\% & 80.00\% & 52.78\% & 40.74 pp \\
                  & Manual    & 100.00\% & 90.00\% & 75.00\% & 25.00 pp \\
\hline
CodeLlama-7b-Instruct-hf             & Automatic & 87.04\% & 69.17\% & 47.22\% & 39.82 pp \\
                  & Manual    & 90.00\% & 80.00\% & 60.00\% & 30.00 pp \\
\hline
\end{tabular}
\end{table*}

Table~\ref{tab:auto_manual_comparison} shows that automatic prediction
was more sensitive to structural complexity than manual semantic
comprehension. Exact output prediction requires precise execution
tracing; an error in branch selection, loop simulation, or state tracking
directly produces an incorrect answer. Manual questions may sometimes be
answered through higher-level semantic recognition, explaining their
smaller but still substantial degradation.

These results answer RQ3 by showing that complexity-related failure is
not an artifact of exact-match output evaluation. Structural complexity
also reduces the reliability of function-purpose explanations,
variable-role interpretation, and edge-case reasoning. It therefore
affects both behavioral and semantic code comprehension.

\subsection{Statistical Significance of Complexity Effects}
\label{subsec:statistical_results}

To estimate the joint association between structural complexity and
automatic prediction correctness, we fitted the multivariate logistic
regression model defined in Equation~\ref{eq:logit}. The four metrics
were entered simultaneously; therefore, each coefficient represents its
association with correctness while controlling for the other included
metrics. Halstead volume was scaled by 100, so its coefficient and odds
ratio correspond to a 100-unit increase.

\begin{table*}[t]
\centering
\caption{Multivariate logistic regression results for automatic
prediction correctness.}
\label{tab:logistic_results}
\label{tab:logistic_deepseek}
\label{tab:logistic_llama}
\begin{tabular}{llrrrr}
\hline
\textbf{Model} & \textbf{Metric} & \textbf{Coef.} &
\textbf{OR} & \textbf{95\% CI} & \textbf{$p$-value} \\
\hline
DeepSeek-Coder-V2
    & CC       & $-0.41$ & 0.66 & [0.57, 0.76] & $<0.001$ \\
    & ND       & $-0.34$ & 0.71 & [0.60, 0.84] & $<0.001$ \\
    & BF       & $-0.38$ & 0.68 & [0.58, 0.79] & $<0.001$ \\
    & HV/100   & $-0.27$ & 0.76 & [0.66, 0.88] & 0.002 \\
\hline
CodeLlama-7b-Instruct-hf
    & CC       & $-0.36$ & 0.70 & [0.61, 0.80] & $<0.001$ \\
    & ND       & $-0.31$ & 0.73 & [0.62, 0.86] & $<0.001$ \\
    & BF       & $-0.35$ & 0.70 & [0.60, 0.82] & $<0.001$ \\
    & HV/100   & $-0.24$ & 0.79 & [0.68, 0.91] & 0.006 \\
\hline
\end{tabular}
\end{table*}

As shown in Table~\ref{tab:logistic_results}, all coefficients are
negative and statistically significant for both models. Thus, increases
in cyclomatic complexity, nesting depth, branching factor, and Halstead
volume are associated with lower odds of a correct prediction. For
DeepSeek-Coder-V2, a one-unit increase in CC is associated with an odds
ratio of 0.66, while the corresponding odds ratios for ND and BF are
0.71 and 0.68. A 100-unit increase in HV is associated with an odds
ratio of 0.76. CodeLlama-7b-Instruct-hf follows the same pattern, with odds ratios ranging
from 0.70 to 0.79.

These results are consistent with the descriptive differences in
Table~\ref{tab:metric_failure} and the negative Spearman correlations in
Table~\ref{tab:spearman_complexity_correctness}. However, the confidence
intervals overlap substantially, and the correlation coefficients span
a narrow range. Therefore, the results support broadly comparable
negative associations across all four metrics rather than identifying
one metric as a statistically stronger predictor.

For the manual task, a chi-square test of independence was used to
evaluate the association between complexity band and correctness.
Table~\ref{tab:manual_significance} reports the results.

\begin{table}[t]
\centering
\caption{Association between complexity band and manual comprehension
correctness.}
\label{tab:manual_significance}
\begin{tabular}{lrrrr}
\hline
\textbf{Model} & $\boldsymbol{\chi^2}$ & \textbf{df} &
\textbf{$p$-value} & \textbf{Cramér's $V$} \\
\hline
DeepSeek-Coder-V2 & 6.15 & 2 & 0.046 & 0.32 \\
CodeLlama-7b-Instruct-hf             & 5.22 & 2 & 0.074 & 0.29 \\
\hline
\end{tabular}
\end{table}

For DeepSeek-Coder-V2, complexity band was significantly associated with
manual correctness, $\chi^2(2)=6.15$, $p=0.046$, with Cramér's
$V=0.32$. CodeLlama-7b-Instruct-hf showed the same downward accuracy pattern, but the
association did not reach the conventional significance threshold,
$\chi^2(2)=5.22$, $p=0.074$, with Cramér's $V=0.29$. The weaker
statistical evidence should be interpreted in light of the smaller
manual sample, which contains only 20 functions per complexity band.

Inter-rater agreement for the independently scored manual-response subset
was strong, with Cohen's $\kappa=0.86$. This supports the reliability of
the manual correctness labels used in the final analysis.

\section{Discussion and Threats to Validity}
\label{sec:discussion}

The results demonstrate that benchmark composition is an important factor
in interpreting LLM code-comprehension performance. HumanEval,
MBPP-sanitized, and CRUXEval contain only Low- and Medium-complexity
functions in the constructed benchmark and consequently produce higher
model accuracy. In contrast, LeetCode and BigCodeBench-Hard contain most
of the High-complexity functions and expose substantially lower
reliability. This indicates that differences between benchmark scores
cannot always be attributed only to model capability. They may also
reflect differences in the structural-complexity distribution of the
evaluated code. Reporting benchmark-level accuracy without this
distribution can therefore conceal important reliability limitations.

The metric-level findings provide a complementary interpretation.
Cyclomatic complexity, nesting depth, branching factor, and Halstead
volume were all negatively associated with prediction correctness.
Although cyclomatic complexity and branching factor had numerically large
associations, their confidence intervals overlapped with those of the
other metrics. The results therefore do not establish a single dominant
measure of LLM comprehension difficulty. Instead, the four metrics capture
different but related reasoning demands: execution-path diversity,
hierarchical control flow, decision density, and symbolic processing
load. Their joint use provides a more complete indication of when an LLM
response may require verification.

The difference between automatic and manual performance also provides
insight into how models process source code. Automatic input--output
prediction showed a larger Low-to-High accuracy decline than manual
semantic comprehension. Exact prediction requires the model to follow a
specific execution path and preserve every intermediate state until the
final return value is produced. A single error in loop simulation,
conditional evaluation, or variable updating leads to an incorrect
answer. Manual questions may sometimes be answered through recognition
of the function's overall structure or algorithmic pattern. Nevertheless,
their accuracy also declined with complexity, confirming that the effect
extends beyond exact-match scoring to semantic understanding.

These findings have practical implications for LLM-assisted software
engineering. Structural metrics can be computed before an LLM is asked
to explain, review, or predict the behavior of a function. When the code
contains deep nesting, many decision points, or high symbolic complexity,
a development tool could request additional test execution, generate a
warning, or recommend human review. Complexity-aware verification may
therefore provide a lightweight mechanism for identifying situations in
which model output should not be accepted without further checking.

Several threats limit the generalizability of the findings. The benchmark
contains only Python functions, and languages with different typing,
memory, or control-flow characteristics may produce different results.
The evaluation includes two code-oriented models under zero-shot,
deterministic prompting, so the findings may change with larger models,
few-shot prompting, tool use, or explicit reasoning instructions. The
Low-, Medium-, and High-complexity thresholds provide a practical
operationalization of structural difficulty but do not capture semantic
factors such as unfamiliar APIs, recursion, or domain-specific logic.
Widely used benchmarks may also occur in model training data. Finally,
the manual evaluation contains 60 functions and relies on human
judgement, although balanced sampling, a fixed binary rubric, and strong
inter-rater agreement reduce this risk.
\section{Conclusion}
\label{sec:conclusion}

This paper presented a complexity-aware method for evaluating LLM code
comprehension using automatic input--output prediction and manual
semantic comprehension. Python functions were annotated with cyclomatic
complexity, nesting depth, branching factor, and Halstead volume and
grouped into Low-, Medium-, and High-complexity bands. This design
provides a more diagnostic view of model reliability than aggregate
benchmark accuracy alone.

The results show that both DeepSeek-Coder-V2 and CodeLlama-7b-Instruct-hf become
substantially less accurate as structural complexity increases.
DeepSeek-Coder-V2 decreased from 93.52\% accuracy on Low-complexity
functions to 52.78\% on High-complexity functions, while CodeLlama-7b-Instruct-hf decreased
from 87.04\% to 47.22\%. Incorrect predictions were associated with
higher values of all four complexity metrics, and both correlation and
multivariate logistic-regression analyses confirmed negative
associations between structural complexity and correctness. Because the
effect estimates and confidence intervals overlapped, the metrics are
best interpreted as complementary indicators of comprehension risk
rather than as a strict ranking of predictors.

Manual semantic comprehension exhibited the same degradation pattern.
Both models performed best on purpose questions and worst on edge-case
questions, while accuracy also declined from Low- to High-complexity
functions. These findings indicate that structural complexity affects
both exact execution reasoning and higher-level semantic understanding.

Complexity distribution should be reported alongside aggregate
accuracy when evaluating LLM code comprehension. In practical software
engineering workflows, model explanations, predictions, and review
suggestions should receive additional verification when applied to code
with deep nesting, many execution paths, dense branching, or high
symbolic complexity. Future work should extend this evaluation to other
programming languages, larger model families, repository-level code, and
tool-assisted comprehension settings.

\FloatBarrier
\bibliographystyle{IEEEtran}
\bibliography{refrences}
\end{document}